\documentclass{llncs}
\usepackage{float}
\usepackage{tabularx,ragged2e}
\usepackage{amsmath}
\usepackage{booktabs}
\usepackage{multirow}
\usepackage{subfigure}
\usepackage{lscape}
\usepackage[table,xcdraw]{xcolor}
\usepackage{makeidx}  
 \usepackage{multirow}
 \usepackage[table,xcdraw]{xcolor}
\usepackage{amssymb}
\usepackage{array}
\newcolumntype{P}[1]{>{\centering\arraybackslash}p{#1}}
\usepackage{graphicx}
\usepackage{tabu}
\usepackage[flushleft]{threeparttable}
\usepackage{url}
\usepackage{algorithm}
\usepackage[noend]{algpseudocode}
\makeindex
\begin{document}

\title{Diabetic Foot Ulcer Grand Challenge 2021: Evaluation and Summary}

\author{Bill Cassidy\inst{1}\orcidID{0000-0003-3741-8120} 
\and Connah Kendrick \inst{1}\orcidID{0000-0002-3623-6598} 
\and Neil D. Reeves \inst{2}\orcidID{0000-0001-9213-4580} 
\and Joseph M. Pappachan \inst{3}\orcidID{0000-0003-0886-5255} 
\and Claire O'Shea \inst{4} 
\and David G. Armstrong \inst{5}\orcidID{0000-0003-1887-9175} 
\and Moi Hoon Yap\inst{1}\orcidID{0000-0001-7681-4287}}
%
\authorrunning{B. Cassidy et al.}
%

\institute{Department of Computing and Mathematics, Manchester Metropolitan University, Manchester M1 5GD, UK \and
Musculoskeletal Science and Sports Medicine, Manchester Metropolitan University, Manchester M1 5GD, UK \and
Lancashire Teaching Hospitals NHS Foundation Trust, Preston, PR2 9HT, UK \and
Waikato District Health Board, Hamilton 3240, New Zealand \and
Southwestern Academic Limb Salvage Alliance (SALSA), Department of Surgery, Keck School of Medicine of University of Southern California, Los Angeles, California, USA\\
\email{B.Cassidy@mmu.ac.uk, M.Yap@mmu.ac.uk}
}

\maketitle


\begin{abstract}
Diabetic foot ulcer classification systems use the presence of wound infection (bacteria present within the wound) and ischaemia (restricted blood supply) as vital clinical indicators for treatment and prediction of wound healing. Studies investigating the use of automated computerised methods of classifying infection and ischaemia within diabetic foot wounds are limited due to a paucity of publicly available datasets and severe data imbalance in those few that exist. The Diabetic Foot Ulcer Challenge 2021 provided participants with a more substantial dataset comprising a total of 15,683 diabetic foot ulcer patches, with 5,955 used for training, 5,734 used for testing and an additional 3,994 unlabelled patches to promote the development of semi-supervised and weakly-supervised deep learning techniques. This paper provides an evaluation of the methods used in the Diabetic Foot Ulcer Challenge 2021, and summarises the results obtained from each network. The best performing network was an ensemble of the results of the top 3 models, with a macro-average F1-score of 0.6307.
\end{abstract}

\section{Introduction}
Diabetic foot ulcers (DFU) are one of the most serious complications that can result from diabetes, and often lead to amputation of all or part of a limb if not met with timely treatment \cite{armstrong2017diabetic,boulton2019diagnosis}. Early detection of DFU, together with accurate screening for infection and ischaemia can help in early treatment and avoidance of further serious complications including amputation. In previous studies, various researchers \cite{wang2015unified,goyal2017fully,goyal2018dfunet,goyal2018robust,yap2020deep} have achieved high accuracy in automated detection of DFUs with machine learning algorithms. A number of widely used clinical DFU classification systems are currently in use, such as Wagner \cite{wagner1987diabetic}, University of Texas \cite{lavery1996classification,armstrong1998validation}, and SINBAD Classification \cite{ince2008use}, which include information on the site of the DFU, area, depth, presence of neuropathy, ischaemia and infection. We focus on ischaemia and infection, which are key features of DFU classification systems and important clinical determinants for effective treatment and healing. This focus is consistent with the evolution of threatened limb classification systems, such as the Wound, Ischemia, and foot Infection (WIfI) classification which is used to predict the risk of amputation in patients diagnosed with critical limb ischemia \cite{mills2014society,armstrong2013juggling}.

Recognition of infection and ischaemia are key determinate factors that predict the healing progress of DFU and risk of amputation. Ischaemia develops due to lack of arterial inflow to the foot, that results in spontaneous necrosis of the most poorly perfused tissues (gangrene), which may ultimately require amputation of part of the foot or leg. In previous studies, it is estimated that patients with critical limb ischaemia have a three-year limb loss rate of approximately 40\% \cite{albers1992assessment}. Patients with an active DFU, particularly those with ischaemia or gangrene, should also be examined for the presence of infection. Approximately, 56\% of DFU become infected and 20\% of DFU infections lead to amputation of foot or limb \cite{prompers2007high,lipsky20122012,lavery2003diabetic}. In one recent study, 785 million patients with diabetes in the US between 2007 and 2013 suggested that DFU and associated infections constitute a powerful risk factor for emergency department visits and hospital admission \cite{skrepnek2017health}. Due to high risks of infection and ischaemia associated with DFU amputation \cite{mills2014society}, timely and accurate recognition of infection and ischaemia in DFU with cost-effective machine learning methods is an important step towards the development of a complete computerised DFU assessment system.

In current practice, DFU assessment is conducted in foot clinics and hospitals by podiatrists and diabetes physicians. To determine appropriate treatment, a vascular assessment is performed for ischaemia and the wound is assessed for clinical evidence of infection and wound tissue sent for microbiological culture. Van Netten et al. \cite{van2017validity} found that clinicians achieved low validity and reliability for remote assessment of DFU in foot images. Hence, it is clear that analysing these conditions from images is extremely difficult even by experienced podiatrists. Patient experiences may be different, however. Swerdlow et al. \cite{swerdlow2021selfie}, instituted a “foot selfie” programme and found overall high levels of patient engagement. Limited research exists using computerised methods to automate the monitoring of DFU using foot photographs \cite{yap2018new}. This is due to the lack of availability of datasets with clinical labelling for research purposes \cite{cassidy2020dfuc}. 

\begin{figure}[!htb]
\centering
\includegraphics[width=1.0\textwidth]{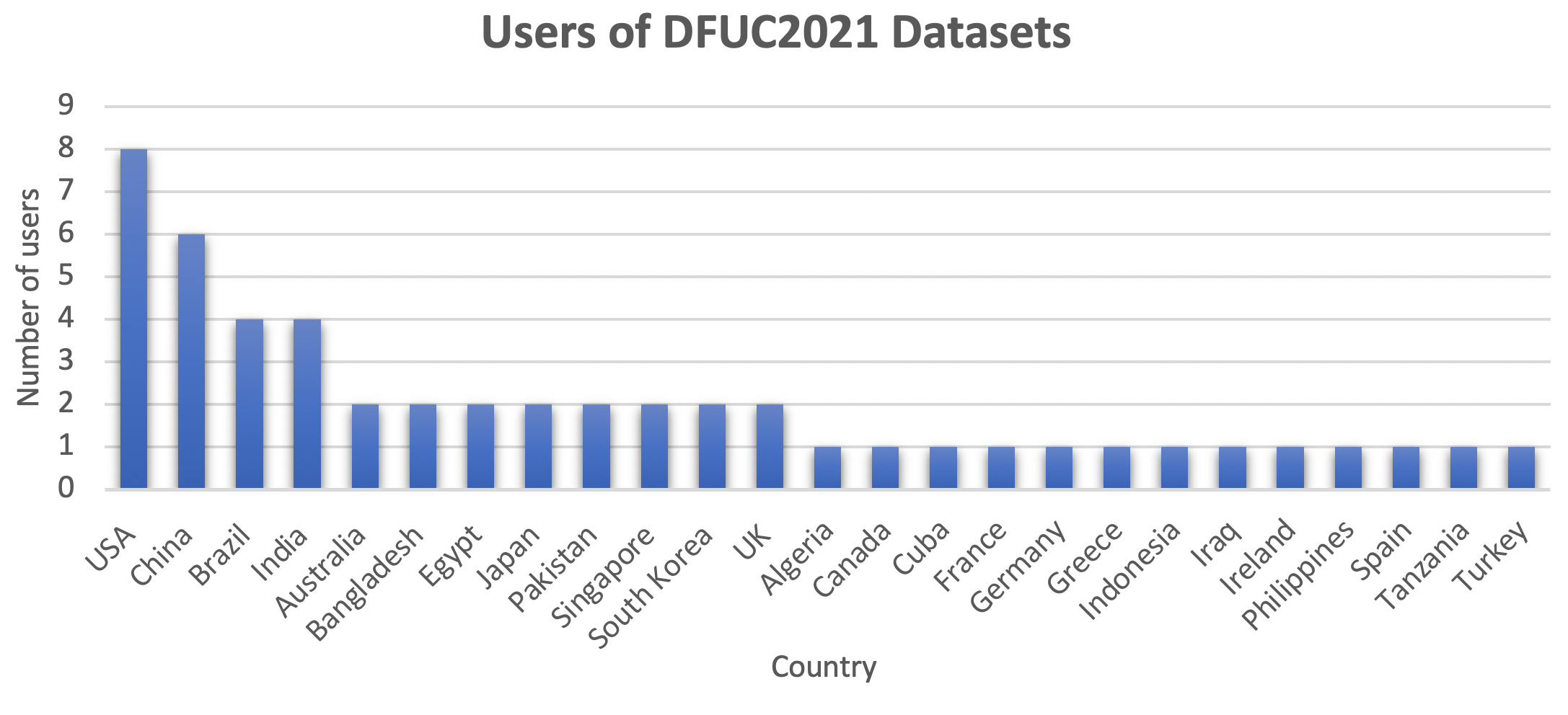}
\caption{Distribution of researchers using the DFUC2021 datasets and their country of origin.}
\label{figure:dfuc2021_users}
\end{figure}

Motivated by technological advancements in medical imaging \cite{esteva2017skin,brinker2019superior,fujisawa2019surpasses,pham2020binary,jinnai2020pigmented}, where machine learning algorithms performed better than experienced clinicians, Goyal et al. \cite{goyal2020recognition} analysed the performance of machine learning algorithms on the recognition of ischaemia and infection on DFUs. They proved that deep learning methods outperformed conventional machine learning methods on a small dataset (1,459 images) and proposed an ensemble Convolutional Neural Network (CNN) approach for ischaemia and infection recognition. Although they achieved high accuracy in ischaemia recognition, there were a number of limitations to their method: 1) the proposed binary classification ensemble CNN method detected one class at a time, which was not capable of detecting co-occurrence of infection and ischaemia; 2) the dataset was small and cannot be generalised; 3) the dataset was highly imbalanced, with infection cases significantly outnumbering ischaemic cases; and 4) the recognition rate of infection was 73\%, which requires substantial work to improve accuracy. To address these issues, Yap et al. \cite{yap2021analysis} introduced the DFUC2021 datasets, which consist of 4,474 clinically annotated images, together with DFU patches with the label of infection, ischaemia, both infection and ischaemia and none of those conditions (control). Since the release of the DFUC2021 datasets on the 15th April 2021, they have been shared with 51 institutions from 25 countries. Figure \ref{figure:dfuc2021_users} illustrates the distribution of researchers using the DFUC2021 datasets by country, showing that the majority of users originate from the United States, China, India and Brazil.

\section{Methodology}
This section summarises the datasets used for DFUC2021, the performance metrics and analysis of the methods proposed by the participants for challenge.

\subsection{Datasets and Ground Truth}
The previous publicly available dataset created by Goyal et al. \cite{goyal2020recognition} consists of 1,459 DFUs: 645 with infection, 24 with ischaemia, 186 with infection and ischaemia, and 604 control DFU (presence of DFU, but without infection or ischaemia). The DFUC2021 dataset is the largest publicly available dataset with DFU pathology labels which consists of 1,703 ulcers with infection, 152 ulcers with ischaemia, 372 ulcers with both conditions, 1703 control DFU and an additional 1,337 unlabelled DFU. The ground truth was produced by two healthcare professionals who specialise in diabetic wounds and ulcers. The instruction for annotation was to label each ulcer with ischaemia, and/or infection, or none. The patient medical record was used to validate the labels. To increase the number of DFU patches for deep learning algorithms, we used natural augmentation \cite{goyal2020recognition} and generated a total of 15,683 DFU patches, which consists of 11,689 labelled patches and 3,994 unlabelled patches. For the labelled DFU patches, the train, validation and test split is: 4,799 patches for the training set, 1,156 patches for the validation set, and 5,734 patches for the testing set. The detailed split for each pathology is presented in Table \ref{tab:split}. As shown in Table \ref{tab:split}, the number of patches for ischaemia, infection and ischaemia are relatively low when compared to the other classes.

\begin{table}[]
	\centering
	\small\addtolength{\tabcolsep}{2pt}
	\renewcommand{\arraystretch}{1}
	\caption{DFUC2021 dataset distribution for training (4,799 patches) and validation (1,156 patches) after natural augmentation.}
	\label{tab:split}
	\scalebox{1.0}{
		\begin{tabular}{cccccc}
			\hline
			 & Infection & Ischaemia & Infection and Ischaemia & None & \textbf{Total}\\ \hline
			Train   &  2074     &   179    &       483           &  2063 & 4799\\ \hline
			Validation     &   481    &       48             & 138  &    489 & 1156\\ \hline
	\end{tabular}}
\end{table}

\begin{figure}
	\centering
	\begin{tabular}{cccc}
		\includegraphics[width=.2\textwidth]{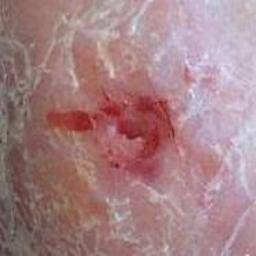} &
		\includegraphics[width=.2\textwidth]{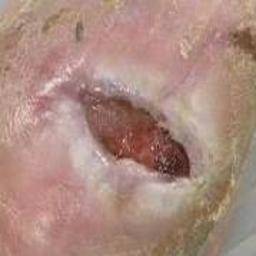} &
		\includegraphics[width=.2\textwidth]{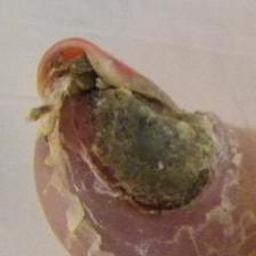} &
		\includegraphics[width=.2\textwidth]{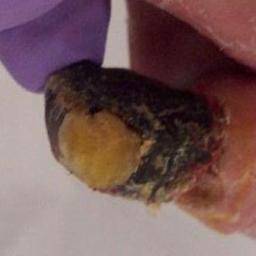}\\
		(a) & (b) & (c) & (d)\\  
	\end{tabular}
	\caption[]{Illustration of DFU patches from the DFUC2021 dataset. From left to right: (a) control DFU, (b) DFU with infection, (c) DFU with ischaemia and (d) DFU with both infection and ischaemia.}
	\label{figure:dataset}
\end{figure}

Figure \ref{figure:dataset} illustrates DFU patches of four conditions. As these ulcers exhibit variability within a single condition and similarity between different conditions, the DFUC2021 dataset presents a significant challenge for computer vision and machine learning methods in the recognition of infection and ischaemia.

\subsection{Performance Metrics} 
We compared the performance of the deep learning networks on recognition of infection and ischaemia using precision, recall, F1-score and area under the Receiver Operating Characteristics Curve (AUC). For the performance in multi-label classification, due to class imbalance inherent within the DFUC2021 dataset, the performance will be reported in macro-average. Macro-average is used in imbalanced multi-class settings as it provides equal emphasis on minority classes \cite{forman2010apples}. To compute macro-average F1-score, first we obtain all the True Positives (TP), False Positives (FP) and False Negatives (FN) for each class $i$ (of n classes), and their respective F1-scores. The Macro-F1 is determined by averaging the per-class F1-scores, $F1_i$:
\begin{align}
    \text{F1}_i =& \frac{2 \cdot TP_i}{2 \cdot TP_i + FP_i + FN_i}\\
    \text{Macro-F1} =& \frac{1}{N}\sum_{i} {F1_i}
\end{align}
where $i={1,..,N}$ represents the $i$-th class and $N$ is the total number of classes, in this case, $N=4$. For completeness, we also compared the performance of the algorithms with macro-average AUC.


\subsection{Analysis of the Proposed Methods}
In this section, we detail the methods used by the top 10 entries for DFUC2021. 

The method ranked 10th in the challenge, submitted by Ye Hai, proposes two classifiers. The first classifier is used to detect control cases, while the second classifier is used to detect the other three categories - infection, ischaemia and both infection and ischaemia. The group tested a variety of classification networks, such as ResNet, ViT, DenseNet and SENet, and found that SENet34 provided the best results for both classifiers.

The method ranked 9th in the challenge, submitted by Weilun Wang, proposes a texture classification model which used SE-DenseNet (Squeeze and Excitation Densely Connected Convolutional Network). SE-DenseNet combines the advantages of DenseNet and SENet, which uses multi-dimensional feature information, strengthening the transmission of deep information and enhancing the learning and expression ability of the deep network through a "feature recalibration'' strategy. Further, the network is also able to slow down the attenuation of errors in each hidden layer, ensuring the stability of gradient weight information and avoiding the disappearance of gradient through the reverse conduction mechanism of the network itself, improving network performance \cite{qu2020deep}. This approach does not require a very deep model, so networks such as DenseNet121 and EfficientNetB0, which contain over 100 convolution layers, were not used. To determine the optimum model depth required for this scenario, Wang performed several experiments. First, the 5,955 training samples were split into 10 subsets, followed by 10-fold cross validation. Next, 9-fold samples were randomly up-sampled and used as training samples in each sub-experiment, with the remaining 1-fold samples used for testing.




%

\begin{figure}
	\centering
	\begin{tabular}{c}
		\includegraphics[width=1\textwidth]{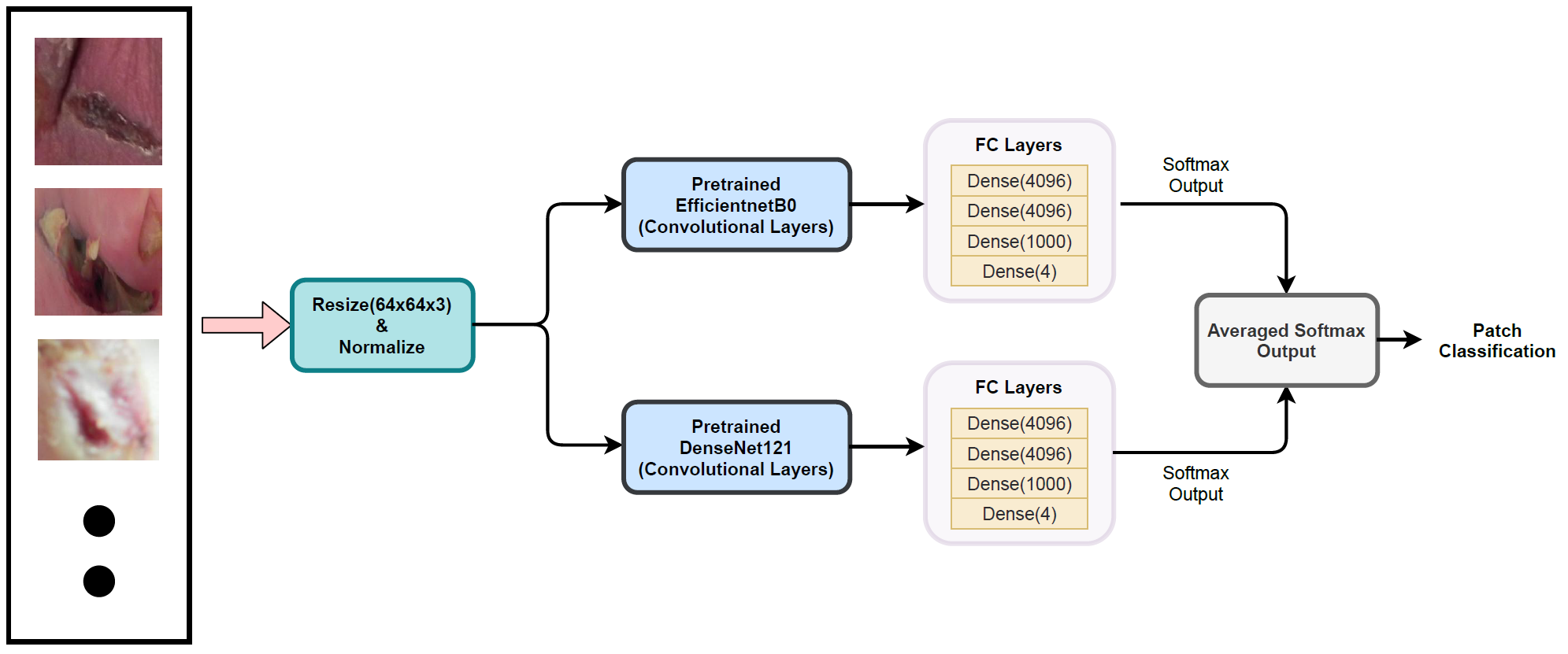}\\
	\end{tabular}
	\caption[]{Illustration of the prediction level ensembling approach used by Das et al., which achieved 8th place in DFUC2021.}
	\label{figure:das_model}
\end{figure}

The method ranked 8th in DFUC2021, submitted by Das et al., proposed a prediction level ensembling. This submission utilised DenseNet121 and EfficientNetB0 models pretrained using ImageNet. The convolutional layers are taken as proposed in the original work, however, the fully connected layers are set as FC(4096), FC(4096), FC(1000) and FC(4), which all use ReLU activations except the final softmax based prediction layer. The configuration of FC layers is motivated by the original VGG16 architecture \cite{simonyan2015deep}. The Softmax predictions from both networks are averaged to obtain a prediction level ensembling, providing a final prediction. Figure \ref{figure:das_model} shows an overview of the network configuration\footnote{reproduced with permission from Sujit Kumar Das, Department of CSE, National Institute of Technology, Silchar 788010, Assam, India}.

The method ranked 7th for DFUC2021, submitted by Chuantao Xie, utlises EfficientNetB0 and DenseNet121, both pretrained using ImageNet. EfficientNetB0 was chosen for its additive feature fusion, while DenseNet121 was chosen for its concatinated feature fusion. The proposed method replaces the layers after the main structure of the CNN, leaving the rest of the network unchanged, including the network structure that proceeds the global average pooling layer which is connected using a parallel structure. One of the branches of the parallel network structure includes convolution, batch normalization, activation function, global average pooling and full connection, followed by the class prediction. Finally, the predicted results of the parallel network structure are concatenated, providing the full connection prediction.



\begin{figure}
	\centering
	\begin{tabular}{c}
		\includegraphics[width=.95\textwidth]{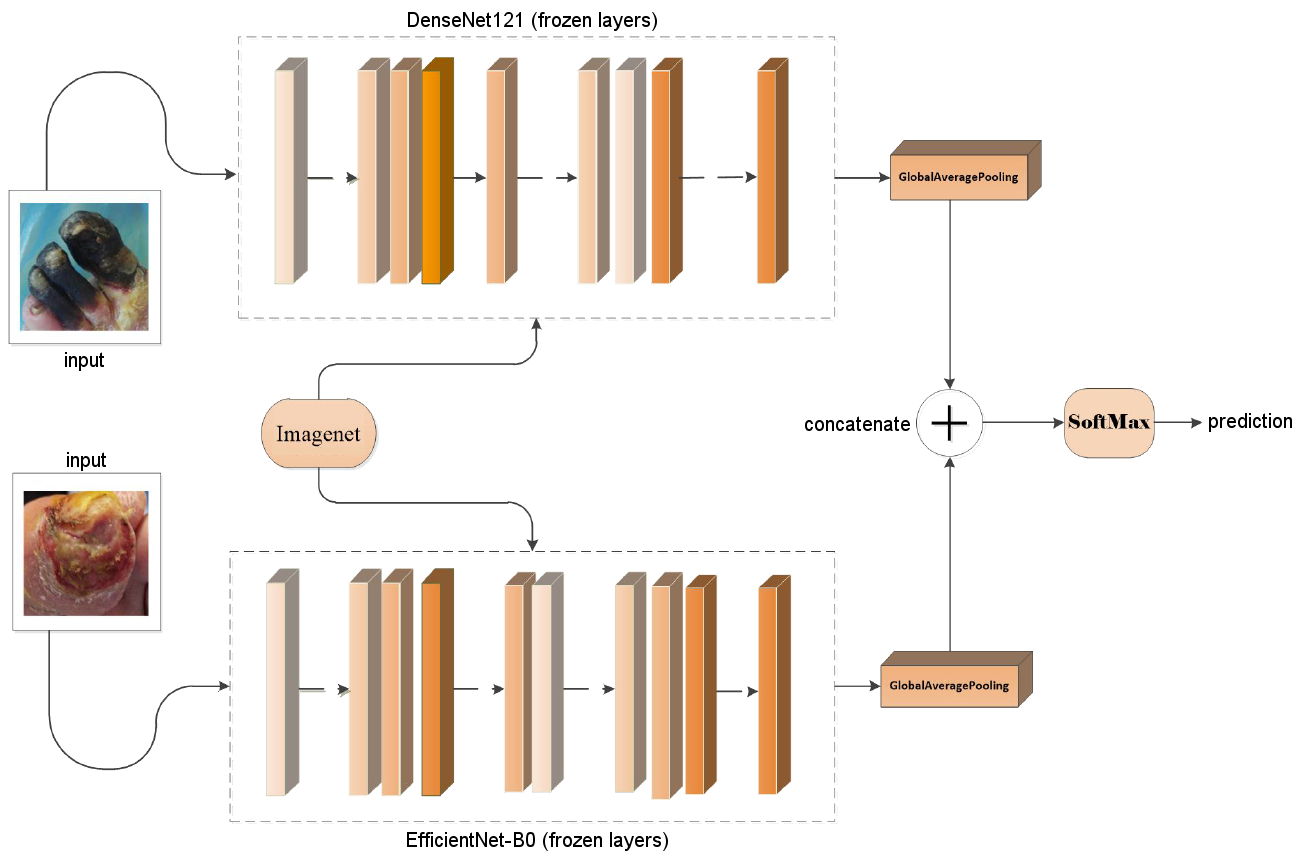}\\
	\end{tabular}
	\caption[]{Illustration of the ensemble CNN approach used by Chen et al., which achieved 6th place in DFUC2021.}
	\label{figure:chen_model}
\end{figure}

The method which placed 6th for DFUC2021, submitted by Chen et al., utilises an ensemble approach using DenseNet121 and EfficientNet pretrained on ImageNet with a frozen output layer connected to a global average pooling layer. Concatenated integration was implemented with two inputs and one output. The fully connected output layer of the pretrained network was replaced by a new four-class SoftMax layer. Figure \ref{figure:chen_model} shows an overview of the network configuration\footnote{reproduced with permission from Donghui Lv, Yuqian Chen, School of Communication and Information Engineering, Shanghai University, Shanghai 200444, China}.

The method ranked at 5th place in the challenge, submitted by Güley et al., leveraged the GenerAlly Nuanced Deep Learning Framework (GaNDLF) to achieve multi-class DFU wound classification. GaNDLF enables various machine learning (ML) and artificial intelligent (AI) workloads, including segmentation, regression, classification, and synthesis. This is achieved using a range of imaging modalities, such as RGB, radiographic and histopathologic imaging techniques. Three VGG architectures were trained (VGG11, VGG16 and VGG19) using the DFUC2021 dataset. VGG was selected due to its use of very small convolutions which utilise spacial padding to preserve features from the input image. A total of 5 max-pooling operations were used over a $2\times N2$ window size, with a stride of 2 to ensure that image dimensions were halved after each max-pooling operation. ReLU activation with global average pooling and two drop-out layers with a penultimate linear layer were used for the classifier. A Softmax layer forms the last layer to provide a final classification. 



The method submitted by Qayyum et al, which ranked 4th in the challenge, utilised Vision Transformers (ViTs) to perform DFU classification. ViTs inherently reduce inductive biases, such as translation variances and locality, which are present in most other CNN architectures. This solution used pretrained ViTs and fine-tuned them on the DFUC2021 dataset. The features obtained from the last network layer from each ViT were concatenated pairwise, followed by a fully connected layer to concatenate features from individual ViTs before being passed to the final classifier. To address the issue of imbalanced class distribution within the DFUC2021 dataset, average weighted sampling was used, which was shown to improve experimental results.



The method submitted by Ahmed et al., which ranked 3rd in DFUC2021, fine-tuned EfficientNet B0-B6, Resnet-50, and Resnet-101 (pretrained on ImageNet) on the DFUC2021 training set and proposed a custom activation layer using Bias Adjustable Softmax. This Softmax-based activation layer is used to handle class imbalance inherent within the dataset. Their initial experiments used weighted categorical cross entropy but found no significant impact on performance. They found that the use of class weights in the loss function resulted in trained networks showing a bias towards control and infection cases. To address this problem, a novel method was introduced to adjust the skew of probabilities for each class to adjust the bias at inference level. 


The method which ranked 2nd place in DFUC2021, submitted by Bloch et al., utilised an ensemble of EfficientNets with a semi-supervised training strategy involving pseudo-labeling for unlabeled images. Their main contribution was the use of Conditional GANs (pix2pixHD) to generate synthetic DFU images to address class imbalance. To achieve this, they created edge masks to indicate regions of interest on the DFUC2021 dataset images.

The winning entry for DFUC2021, submitted by Galdran et al., compared established CNNs (ResNeXt50 and EfficientNet-B3 pretrained on ImageNet) with a Vision Transformer (ViT) and Data-efficient Image Transformers (DeiT) for DFU multi-classification. They also demonstrated how the Sharpness-Aware Minimization (SAM) optimisation algorithm significantly improves the generalisation capability of both traditional CNNs and ViTs in this domain compared to standard Stochastic Gradient Descent (SGD). SAM seeks parameters that lie in neighbourhoods that have uniformly low loss, which results in a min-max optimisation problem on which gradient descent can be performed efficiently, as shown in Algorithm \ref{algorithm:sam}. This method was developed to address the problem of heavily overparameterised models where training loss values do not always reflect how well the model generalises. SAM has also been shown to improve robustness against label noise \cite{foret2021sam}. Their winning entry utilised a linear combination of predictions extracted from BiT-ResNeXt50 (derived from Big Image Transfer) and EfficientNet-B3 models trained on different data folds. This winning submission achieved the highest F1-score (62.16), AUC (88.55) and recall (65.22) measures for DFUC2021.

\begin{algorithm}
\caption{Pseudocode for the SAM algorithm used by the winning entry for DFUC2021, originally proposed by Foret et al. \cite{foret2021sam}.}
\label{algorithm:sam}
\hspace*{\algorithmicindent} \textbf{Input} Training set, loss function, batch size, step size, neighborhood size. \\
\hspace*{\algorithmicindent} \textbf{Output} Model trained with SAM.
\begin{algorithmic}[1]
\State Initialise weights $w_0$, $t = 0$
\While{\textit{not converged}}
    \State Sample batch $\beta = \{(x_1,y_1),...(x_b,y_b)\}$
    \State Compute gradient $\nabla_wL_\beta(w)$ of the batch’s training loss
    \State Compute $\hat{\epsilon}(w)$
    \State Compute gradient approximation for the SAM objective: $g =  \nabla_wL_\beta(w)|_{w+\hat{\epsilon}(w)}$
    \State Update weights: $w_{t+1} = w_t - \eta{g}$
    \State $t = t + 1$
\EndWhile
\State return $w_t$
\end{algorithmic}
\end{algorithm}

\section{Results and Discussion}
For DFUC2021, there were 500 submissions for the validation stage and 28 submissions for the testing stage. Table \ref{table:test_results} shows a summary of the top 10 best performing networks submitted to the DFUC2021 test leaderboard.

\subsection{Analysis on the Top-3 Results}
In this section, we conduct a statistical analysis of the top-3 results from DFUC2021. Galdran et al. achieved the best F1-score for detection of control cases (0.76), which was an improvement of 0.03 on the baseline (0.73). Ahmed et al. achieved the best F1-score for infection classification (0.68), an improvement of 0.12 on the baseline (0.56). Bloch et al. achieved the best F1-score for ischaemia classification (0.56) which shows an improvement of 0.12 on the baseline (0.44). For F1-score of detection of both infection and ischaemia, Galdran et al. achieved a value of 0.56, resulting in an improvement of 0.09 over the baseline (0.10).

For the micro-average results, Galdran et al. achieved the highest micro-average F1-score (0.68), which is an improvement of 0.05 over the baseline (0.63). They also achieved the highest AUC (0.91), an improvement of 0.04 on the baseline result (0.87).

For the macro-average results, Bloch et al. achieved the highest macro-average precision (0.62) which is an increase of 0.05 over the baseline (0.57). Galdran et al. achieved the best micro-average recall result (0.66), demonstrating an increase of 0.04 over the baseline (0.62). Galdran et al. achieved the highest macro-average F1-score (0.62), which represents an increase of 0.07 over the baseline (0.55). For macro-average AUC, Galdran et al. achieved the best result with (0.89), which is an increase of 0.03 over the baseline result (0.86).

To summarise, the top 3 highest performing entries came from submissions by Galdran et al., Bloch et al. and Ahmed et al. Galdran et al. achieved the highest F1-score for control (0.76), infection and ischaemia (0.57), micro-average F1-score (0.68), micro-average AUC (0.91), and highest macro-average recall (0.66), macro-average F1-score (0.62) and macro-average AUC (0.89). Bloch et al. achieved the highest F1-score for ischaemia classification (0.56), and macro-average precision (0.62). Ahmed et al. achieved the highest F1-score for infection classification (0.68).

The results from the challenge represent a modest increase in performance metrics when compared to the baseline results (mean = 0.064, standard deviation = 0.035, error = 0.011). Possible reasons for this include significant class imbalance inherent within the dataset and the small size of the sample images. F1-scores for infection cases (0.68) and ischaemic cases (0.56) are significantly lower than control cases (0.76), which is a possible further reflection of the class imbalance within the dataset. Additional curation together with additional inter- and intra-rater reliability measures may help to further enhance the datasets. However, dataset curation is a difficult and time-consuming task, and presents additional challenges in the form of label noise and artefacts which could affect the true accuracy of models trained on our data \cite{wen2021datasets,cassidy2021isic}.

\begin{table*}
	\centering
	\renewcommand{\arraystretch}{1.0}
	\caption{Summary of the top 10 performing networks for DFUC2021, compared to the baseline result. AP = Abdul-prediction, AE = Adrian-ensemble, AR = Adrian-results, AM = Ahmed-moded, LE = Louise-ensemble, LID = Louise-ID, ST = shimmer-test, XT = xie-test.}
	\scalebox{0.83}{
	    \label{table:test_results}
		\begin{tabular}{|l|cccc|cc|cccc|}
			\hline
			Method  &\multicolumn{10}{l|}{Metrics} \\
	               & \multicolumn{4}{l}{Per class F1-score} & \multicolumn{2}{l}{micro-average} & \multicolumn{4}{l|}{macro-average}\\
			
		  &Control & Infection & Ischaemia & Both & F1 & AUC & Precision & Recall & F1 & AUC\\
			\hline
			\hline
			Baseline \cite{yap2021analysis} &0.73 &0.56 & 0.44 & 0.47 & 0.63 & 0.87& 0.57 & 0.62 & 0.55&0.86\\ \hline
			
AP\_vit\_bas\_GP4EVbn&0.74&0.61&0.43&0.33&0.63&0.87&0.52&0.59&0.53&0.85 \\
AP\_vit\_mil\_UNKBe8A&0.73&0.55&0.52&0.42&0.62&0.87&0.57&0.59&0.56&0.84 \\
AP\_vit\_multi1\_test&0.75&0.63&0.47&0.43&0.66&0.87&0.58&0.61&0.57&0.85 \\ \hline

AE\_bit\_effb3\_F2&\textbf{0.76}&0.64&0.53&0.56&\textbf{0.68}&\textbf{0.91}&0.61&0.65&\textbf{0.62}&\textbf{0.89} \\
AE\_results\_final\_test&0.74&0.61&0.49&0.49&0.65&0.90&0.58&0.61&0.58&0.88 \\
AE\_results\_final\_test2&\textbf{0.76}&0.64&0.51&0.54&0.68&0.90&0.61&0.65&0.61&0.88 \\

AR\_final\_test4&0.73&0.63&0.52&0.50&0.66&0.90&0.59&0.62&0.60&0.87 \\
AR\_final\_test5&0.75&0.63&0.51&\textbf{0.57}&0.67&0.90&0.61&\textbf{0.66}&\textbf{0.62}&0.88 \\ \hline

AM\_v0\_89\_test&0.72&0.67&0.46&0.54&0.67&0.89&0.60&0.60&0.60&0.86 \\
AM\_v0\_89\_test\_1&0.71&\textbf{0.68}&0.46&0.53&0.67&0.89&0.60&0.60&0.60&0.86  \\ \hline

Arnab&0.73&0.57&0.40&0.45&0.62&0.88&0.53&0.57&0.54&0.85 \\ \hline

LE\_predictio\_aCYsozF&0.75&0.59&\textbf{0.56}&0.54&0.65&0.87&\textbf{0.62}&0.62&0.61&0.86 \\
LE\_Predictio\_SSufTEW&0.74&0.60&0.52&0.52&0.65&0.89&0.60&0.62&0.59&0.87 \\

LID47\_predictions&0.74&0.58&0.54&0.51&0.65&0.87&0.61&0.61&0.60&0.85 \\
LID48\_predictions&0.74&0.59&0.55&0.51&0.65&0.85&0.61&0.62&0.60&0.84  \\
LID49\_predictions&0.74&0.59&0.54&0.53&0.65&0.87&0.61&0.63&0.60&0.86\\ \hline

Orhun&0.74&0.55&0.52&0.44&0.62&0.89&0.59&0.58&0.56&0.87 \\ \hline

ST\_submit1&0.73&0.55&0.48&0.41&0.61&0.87&0.57&0.60&0.54&0.86 \\
ST\_submit2&0.74&0.60&0.45&0.43&0.64&0.88&0.56&0.59&0.55&0.86 \\
ST\_submit3&0.74&0.57&0.49&0.38&0.63&0.87&0.57&0.59&0.55&0.86 \\
ST\_submit4&0.73&0.57&0.46&0.46&0.63&0.87&0.57&0.58&0.56&0.86 \\ \hline

Weilunwang&0.70&0.54&0.42&0.47&0.60&0.86&0.54&0.57&0.53&0.82 \\ \hline
Yeah&0.72&0.55&0.47&0.35&0.60&0.74&0.53&0.56&0.52&0.70 \\ \hline

xie-s\_9163&0.72&0.52&0.50&0.46&0.60&0.87&0.57&0.58&0.55&0.86 \\
XT\_eff\_dense\_91\_I0yNTTP&0.74&0.56&0.46&0.46&0.63&0.88&0.57&0.60&0.55&0.87 \\
			\hline
		\end{tabular}
   }
\end{table*}

\subsection{Ensemble of the Top 10 Performing Models}
In this section, we analyse the results of ensembling the top performing models submitted to DFUC2021 to determine if an ensemble approach can provide an increase to performance metrics in multi-classification of DFU. Table \ref{table:ensemble} shows the results of ensembling the top 10 performing models from DFUC2021.

\begin{table*}
	\centering
	\renewcommand{\arraystretch}{1.0}
	\caption{Summary of the results for the top 10 teams and further analysis on the ensembled results. Ensemble Top 2 represents an ensemble of the top 2 teams results, Ensemble Top 3 represents an ensemble of the top 3 teams results, etc.}
	\scalebox{0.83}{
	    \label{table:ensemble}
		\begin{tabular}{|l|cccc|cc|cccc|}
			\hline
			Method  &\multicolumn{10}{l|}{Metrics} \\
	               & \multicolumn{4}{l}{Per class F1-score} & \multicolumn{2}{l}{micro-average} & \multicolumn{4}{l|}{macro-average}\\
			
		  &Control & Infection & Ischaemia & Both & F1 & AUC & Precision & Recall & F1 & AUC\\
			\hline
		  Top-1&0.7574&0.6388&0.5282&0.5619&0.6801&0.9071&0.6140&0.6522&0.6216&0.8855 \\ \hline
Top-2&0.7453&0.5917&0.5580&0.5359&0.6532&0.8734&0.6207&0.6246&0.6077&0.8616 \\ \hline
Top-3&0.7157&0.6714&0.4574&0.5390&0.6714&0.8935&0.5984&0.5979&0.5959&0.8644 \\ \hline
Top-4&0.7466&0.6281&0.4670&0.4347&0.6577&0.8731&0.5814&0.6104&0.5691&0.8488 \\ \hline
Top-5&0.7360&0.5468&0.5216&0.4396&0.6199&0.8865&0.5917&0.5759&0.5610&0.8702 \\ \hline
Top-6&0.7320&0.5732&0.4621&0.4599&0.6292&0.8725&0.5692&0.5823&0.5568&0.8635 \\ \hline
Top-7&0.7407&0.5566&0.4602&0.4558&0.6253&0.8821&0.5705&0.6032&0.5533&0.8698 \\ \hline
Top-8&0.7275&0.5701&0.4000&0.4463&0.6222&0.8821&0.5329&0.5692&0.5360&0.8471 \\ \hline
Top-9&0.6996&0.5412&0.4237&0.4657&0.5999&0.8622&0.5371&0.5681&0.5326&0.8222 \\ \hline
Top-10&0.7192&0.5456&0.4656&0.3532&0.6027&0.7443&0.5300&0.5611&0.5209&0.7020 \\ \hline \hline
Ensemble Top 2&0.7455&0.6014&0.5615&0.5301&0.6572&0.9054&0.6187&0.6297&0.6096&0.8866 \\ \hline
Ensemble Top 3&0.7491&0.6303&0.5637&0.5799&0.6756&\textbf{0.9096}&\textbf{0.6352}&\textbf{0.6422}&\textbf{0.6307}&0.8870 \\ \hline
Ensemble Top 4&0.7578&0.6410&0.5513&0.5412&0.6805&0.9102&0.6244&0.6416&0.6228&0.8893 \\ \hline
Ensemble Top 5&0.7571&0.6337&0.5653&0.5411&0.6775&0.9112&0.6314&0.6395&0.6243&0.8933 \\ \hline
Ensemble Top 6&0.7566&0.6287&0.5437&0.5383&0.6740&0.9104&0.6253&0.6323&0.6168&0.8947\\ \hline
Ensemble Top 7&0.7611&0.6244&0.5417&0.5137&0.6720&0.9099&0.6201&0.6290&0.6102&0.8968\\ \hline
Ensemble Top 8&0.7618&0.6240&0.5486&0.5309&0.6738&0.9106&0.6251&0.6357&0.6163&0.8980\\ \hline
Ensemble Top 9&0.7615&0.6215&0.5629&0.5292&0.6730&0.9093&0.6291&0.6396&0.6188&0.8964\\ \hline
Ensemble Top 10&0.7628&0.6242&0.5603&0.5209&0.6740&0.9082&0.6280&0.6381&0.6171&0.8954\\ \hline
		\end{tabular}
   }
\end{table*}


\subsection{Visual Comparison of the Top-10 Methods}

We conducted further analysis on the top performing methods to determine trends in the data for images that were both easily predicted correctly with high confidence and images where correct classification was difficult. We then visualised those images and identify key features that could have effected the classification result.

\begin{table}[]
	\centering
	\small\addtolength{\tabcolsep}{2pt}
	\renewcommand{\arraystretch}{1}
	\caption{Images from the testing set which the top 10 networks all predicted correctly.}
	\label{table:ResultsTop10Correct}
	\scalebox{1.0}{
		\begin{tabular}{cccc}
			\hline
			Class & Image 1 & Image 2 & Image 3 \\ \hline
			None & \includegraphics[width=.2\textwidth]{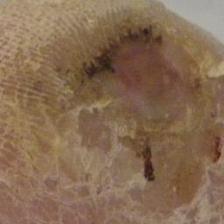} & 
			    \includegraphics[width=.2\textwidth]{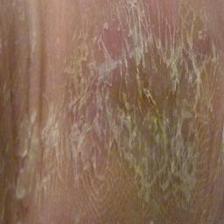} & 
			    \includegraphics[width=.2\textwidth]{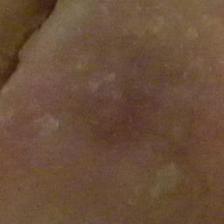} \\ \hline
			    
			Infection & \includegraphics[width=.2\textwidth]{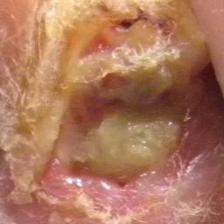} & 
			    \includegraphics[width=.2\textwidth]{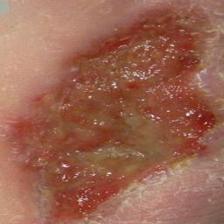} & 
			    \includegraphics[width=.2\textwidth]{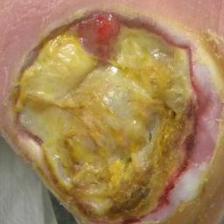} \\ \hline
			
			Ischaemia & \includegraphics[width=.2\textwidth]{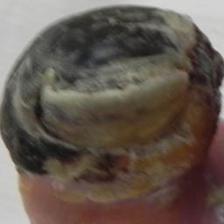} &
			    \includegraphics[width=.2\textwidth]{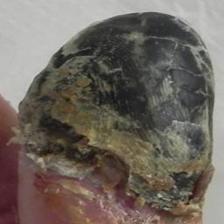} & 
			    \includegraphics[width=.2\textwidth]{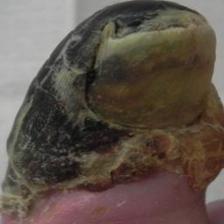}\\ \hline

			Both & \includegraphics[width=.2\textwidth]{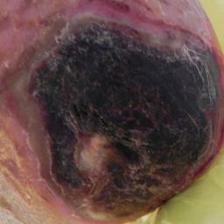} &
			    \includegraphics[width=.2\textwidth]{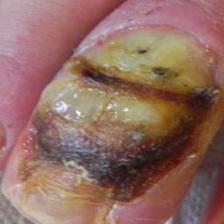} & 
			    \includegraphics[width=.2\textwidth]{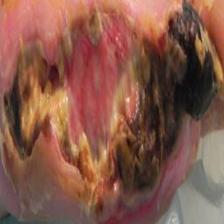} \\ \hline
			
	\end{tabular}}
\end{table}

Table \ref{table:ResultsTop10Correct} highlights 3 images from each class that where correctly classified with high confidence by the top 10 challenge participants. The images show that when the wound is fully visible, the networks are able to determine the difference between classes, even when features from other classes are present, e.g. None image 1 where the black section could cause Ischaemia bias. In-contrast \ref{table:ResultsTop10Incorrect} show examples of images that were incorrectly classified by all top 10 challenge participants. These examples highlight the issue with extreme angles and image blur, as seen in images 1 and 2 for the None class, and in image 2 for Both.

One particularly notable result can be seen when comparing the correctly classified image in Table \ref{table:ResultsTop10Correct} (Image 1, Both) with the incorrectly classified image in Table \ref{table:ResultsTop10Incorrect} (Image 1, Both). The image in Table \ref{table:ResultsTop10Incorrect} is the result of subtle natural augmentation and has resulted in an incorrect classification.


\begin{table}[]
	\centering
	\small\addtolength{\tabcolsep}{2pt}
	\renewcommand{\arraystretch}{1}
	\caption{Images from the testing set which the top 10 networks all predicted incorrectly.}
	\label{table:ResultsTop10Incorrect}
	\scalebox{1.0}{
		\begin{tabular}{cccc}
			\hline
			Class & Image 1 & Image 2 & Image 3 \\ \hline
			None & \includegraphics[width=.2\textwidth]{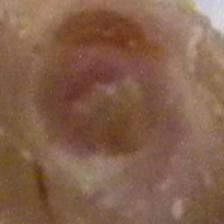} & 
			    \includegraphics[width=.2\textwidth]{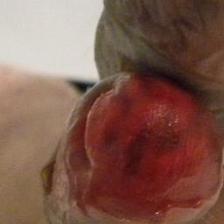} & 
			    \includegraphics[width=.2\textwidth]{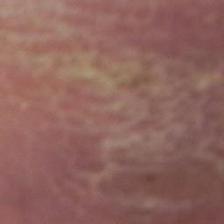} \\ \hline
			    
			Infection & \includegraphics[width=.2\textwidth]{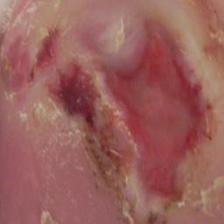} & 
			    \includegraphics[width=.2\textwidth]{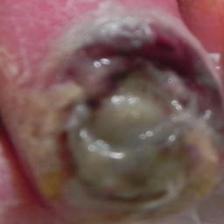} & 
			    \includegraphics[width=.2\textwidth]{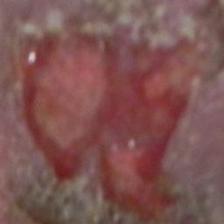} \\ \hline
			
			Ischaemia & \includegraphics[width=.2\textwidth]{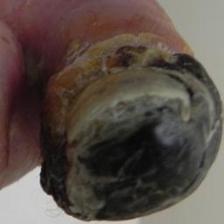} & 
			    \includegraphics[width=.2\textwidth]{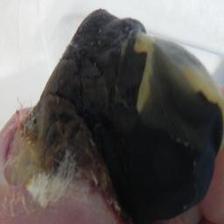} & 
			    \includegraphics[width=.2\textwidth]{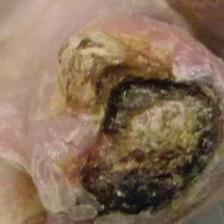}\\ \hline
			
			Both & \includegraphics[width=.2\textwidth]{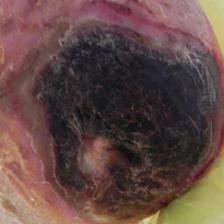} & 
			    \includegraphics[width=.2\textwidth]{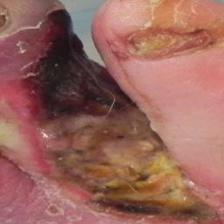} &
			    \includegraphics[width=.2\textwidth]{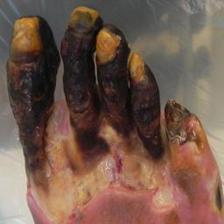} \\ \hline
			
	\end{tabular}}
\end{table}

\section{Conclusion}
In this study, we introduce the largest DFU pathology dataset, and propose a weakly supervised framework for DFU pathology classification of infection and ischaemia. This is the first dataset of its kind to be made available to the research community together with implementation of CNNs for multi-class classification of infection, ischaemia, and co-occurrences of infection and ischaemia. These advancements will help to support early identification of DFU complications to guide treatment and help prevent further complications including limb amputation.

Although the majority of the deep learning methods reported in this paper show promising results in recognising infection and ischaemia, there are still significant challenges in designing methods to detect the co-occurrences of both conditions. Future work will investigate more advanced techniques such as generative adversarial networks and unsupervised learning to improve network performance.

This work will form an important contribution to our ongoing research into developing a fully automated DFU diagnosis and monitoring framework which can be used by patients and their carers in home settings, to help reduce strains on healthcare services around the world. This work will build on our existing framework \cite{reeves2021diabetes,cassidy2021cloudbased} in delivering an easy-to-use system capable of advanced forms of diabetic foot analysis, which will include longitudinal monitoring as a means of assessing wound healing progress.

\section*{Acknowledgment}
We gratefully acknowledge the support of NVIDIA Corporation who provided access to GPU resources for the DFUC2021 Challenge and an NVIDIA Geforce RTX 3090 GPU card as the prize for the winning team.

\addtocmark[2]{Author Index} 
\renewcommand{\indexname}{Author Index}
\printindex

\bibliographystyle{unsrt}
\bibliography{Ref}

\end{document}